\documentclass[ amsmath,amssymb, aps, prl,twocolumn,groupedaddress,cite]{revtex4-1}
\usepackage{graphicx}
\usepackage{dcolumn}
\usepackage{bm}
\usepackage{units}
\usepackage{color}
\usepackage{natbib}
\usepackage{textcomp}
\usepackage{siunitx}
\usepackage{ upgreek }

\begin{document}

\title{Vicinal metal surfaces as potential catalysts for phosphorene epitaxial growth}

\author{Daniel Hashemi}
\email{daniel.hashemi.ctr@us.af.mil}
\author{Gene Siegel}
\author{Michael Snure}
\author{Stefan C. Badescu}
\affiliation{Air Force Research Laboratory, WPAFB, Ohio 45433, USA}


\date{\today}

\begin{abstract}
Phosphorene, a single layer of black phosphorous (BLK-P), has a significant potential for flexible and tunable electronics, but attempts to grow it epitaxially have been unsuccessful to date. Meanwhile, hexagonal blue phoshorous (BL-P) has been achieved on closed-packed $(111)$ metal surfaces in special growth conditions of high vapor pressure and high reactivity of phosphorous. The  $(111)$ surfaces favors BL-P over BLK-P due to its hexagonal symmetry. Here, we investigate computationally the alternative offered by stepped substrates. Using the Cu$(311)$ surface as a model, we find that surface steps can favor energetically BLK-P over BL-P. This can be rationalized in terms of surface density of states and orbital hybridization, which lead to a stronger surface bonding of the lower BLK-P half-layer. This work suggests that vicinal metal surfaces of metals can offer a viable path towards phosphorene synthesis.
\end{abstract}

\maketitle


Mono or few-layer two-dimensional (2D) materials are extremely flexible, have atomically abrupt interfaces, and in principle can be assembled into 3D structures without typical lattice matching requirements characteristic to covalent epitaxy. The few-layer materials are held together by weak van der Waals (vdW) forces. In recent years BLK-P has emerged theoretically as a promising quasi (2D) electronic material characterised by high mobility, ambipolar transport and tunable bandgap \cite{Liu2014}. Bulk black phosphorous is available for experiments, which inspired exfoliation of BLK-P by analogy to graphene exfoliated from bulk graphite. Exfoliated multiple-layer BLK-P flakes has remained the only viable method to obtain BLK-P. In contrast to other 2D materials which were grown on metal surfaces  ({\it e.g.}, graphene \cite{Loginova2009}, hexagonal boron nitride (hBN) \cite{Siegel2017}, silicene \cite{Vogt2012}, germanene \cite{Cantero2017}), the epitaxial growth of BLK-P was not succesful to date, limiting its lateral size and alignment with various substrates. While the growth of 2D materials mentioned above on $(111)$ surfaces was enabled by their $C_{3v}$ symmetry, the $C_{2v}$ symmetry and strong in-plane anisotropy of BLK-P preclude the growth of BLK-P on these substrates. Instead, the theoretically predicted hexagonal BL-P \cite{BL-P_theory} has been obtained \cite{blue-P_growth}.

We attempted to develop a thin film growth epitaxiual method for BLK-P by using the chemical vapor deposition (CVD) applied previously with success to the growth of (layered) hBN \cite{Siegel2017}. We used either phosphane PH$_{3}$ or tributyl phosphate TBP precursors. Even at temperatures as low as $300$\textdegree{}C on copper surfaces we were able to obtain only metal phosphides (P dissolved in the substrate). This is inferred from X-ray diffraction and Raman characterisation, which show a very clear structural change from polycrystalline Cu to Cu$_{3}$P. Similarly, we have observed formation of phosphides on Ni, Au, and Ag at low temperatures. This highlights the high reactivity of P precursor fragments under CVD conditions. It confirms the conclusion of modeling efforts suggesting that flakes of BLK-P are not structurally stable on low-index transition metal surfaces \cite{Gao2016}. It may be possible that the growth of BLK-P be achieved by molecular beam epitaxy (MBE), as in the case of BL-P. For that, an appropriate metal substrate needs to be found that would favor the formation of BLK-P over BL-P.  Several routes can be followed to  alter the catalytic activity of metals, including the demonstration of hybrid metal-hBN substrate with an ‘adjusted van-der Waals bonding’ that could stabilize the BLK-P \cite{Gao2016}, tailoring the chemical activity of the substrate by alloying, or using stepped surfaces \cite{Davidson1994} with lower symmetries and increased chemical activity in one direction. In this study, we follow the latter route.

\begin{figure}[th!]
\centering
\includegraphics[trim = 0cm 0.0cm 0.0cm 0.0cm, clip, scale=0.30, angle=0]{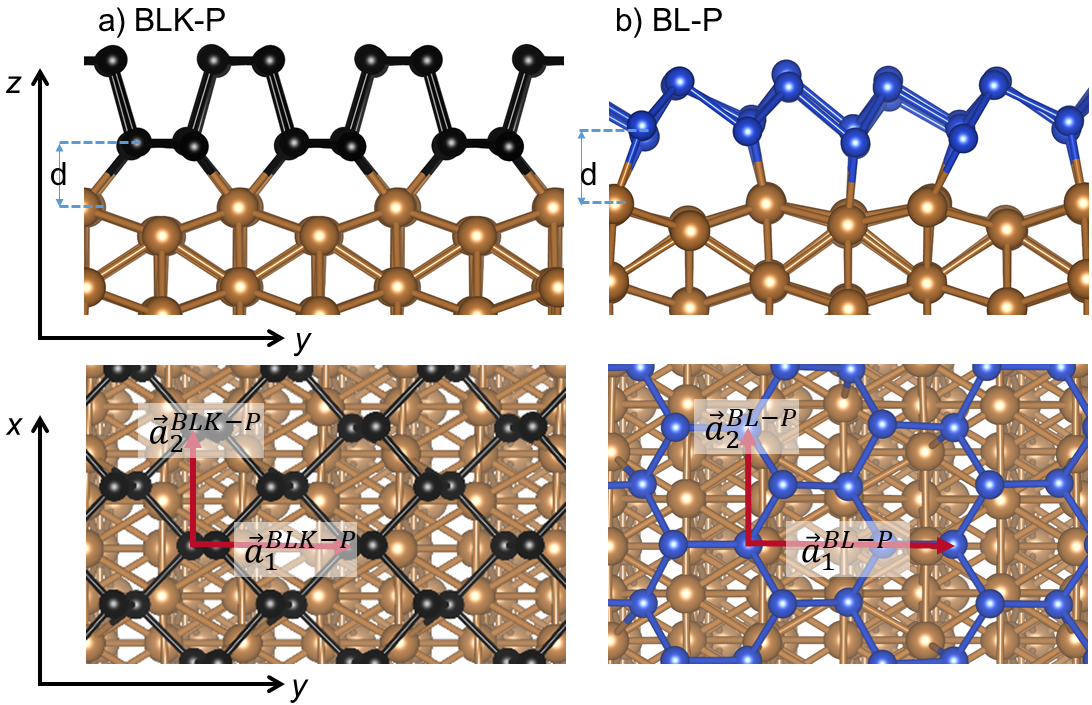}
\caption{ \label{fig:1} Lowest energy structures from first-principles modeling in this work: top and side views of (a) BLK-P and (b) BL-P on the Cu$(311)$ surface. The spacings ($d$) between adsorbed BLK-P, BL-P systems and metal substrate are $1.68$~\AA~and $1.90$~\AA, respectively. The lattice constants of BLK-P and BL-P are highlighted by the red vectors on the lower panel.}
\end{figure}

In this paper, we investigate computationally the possibility raised that stepped surfaces catalyse the growth of BLK-P. This approach is in the spirit of studies of BLK-P vs BL-P energetics on low index metals or semiconductors, which indicated that low-index surfaces favor BL-P over BLK-P, with the exception of a theoretical Sn surface \cite{PRLZeng2017,Qiu2018}.  To study the increased catalytic strength induced by surface step we used Cu since its low-index surface has a lower catalytic activity by comparison to, {\it e.g}., Ni, Ir or Fe \cite{Bokdam2014}. We chose the Cu$(311)$ terrace as a minimal system size for periodic cell calculations that has also a relatively small lattice mismatch with BLK-P compared to other metals.

The calculations were performed using the Vienna \emph{ab initio} simulation package (VASP) \cite{vasp1,vasp2} and based on frozen-core full-potential projector augmented-wave method (PAW) \cite{paw1} with the Perdew, Burke and Ernzernhof \cite{PBE1996} generalized gradient approximation. The convergence with respect to the size of $k$-sampling, cutoff radius, substrate and vacuum thickness was checked. These resulted in a $k$-point mesh of 8$\times$8$\times$1 for density of states (DOS), a plane-wave cutoff radius of $400$~eV, a substrate thickness of six Cu layers, and a vacuum distance between the periodic images of $20$~\AA. Dispersion corrections to standard density functional theory must be included in modeling surface chemistry of 2D materials. Among the vdW -corrected functionals, Grimme's empirical long range dispersive force correction (D2) \cite{Grimme2006} has been able to qualitatively capture the charge redistribution and interlayer spacing of BLK-P system \cite{Shulenburger2015} and has been shown to be accurate for predicting the energetics of rare gas atoms on metal surfaces \cite{Chen2011,Chen2012}. In this work, we used this correction during structural relaxation. We used the conjugate-gradient algorithm for structural relaxation with a force convergence criterion of $0.01$~eV/\AA. For the relaxed structures, we used also the self-consistent method of Tkatchenkoand Scheffler (TS) \cite{Tkatchenko2009} {\it a posteriori} to verify the ordering of converged energies.

The calculated lattice constants of freestanding BLK-P are $a_{1(0)}^{BLK-P}$$=$$4.59$~\AA~and $a_{2(0)}^{BLK-P}$$=$$3.30$~\AA~(armchair and zigzag directions, respectively). For freestanding BL-P, the lattice constant is $a^{BL-P}$$=$$3.27$~\AA; for strain reference we use the zigzag and armchair lattice vector parameters $a_{1(0)}^{BL-P}$$=$$5.66$~\AA~and $a_{2(0)}^{BL-P}$$=$$3.27$~\AA~ (armchair and zigzag directions, Fig.\ref{fig:1}). The bulk metal lattice constant was estimated to be $a^{Cu}$$=$$3.63$~\AA. The corresponding lattice mismatches along the principal directions of BLK-P or BL-P are listed in Tab. \ref{tab:1}. The standard dipole correction for surface adsorption calculations was used to guarantee convergence. The clean metal slab was first relaxed with the three bottom layers fixed at their ideal bulk positions. The top three metal layers were further relaxed again in the presence of BLK-P or BL-P.  For structural relaxations of the adsorbed system in the large supercells, we used only the $\Gamma$ point sampling of the Brillouin zone.

\begin{table}[h!]
	\caption{\label{tab:1} Lattice mismatches $\delta$ and calculated formation energy $E_f$ (Eq.\ref{equ:Ef}) and adsorption energy  $E_{ads}$ (Eq.\ref{equ:Eads}) of BL-P and BLK-P on Cu$(111)$ and Cu$(311)$ substrates. The subscripts TS and D2 indicate the vdW correction scheme. $\delta_{1}$ and $\delta_{2}$ measure lattice mismatches along $y$ and $x$ directions, respectively. A negative value of  $\delta$ indicates that the overlayer is compressively stressed. All energy values are given in meV.}
	\begin{ruledtabular}
		\begin{tabular}{cccc}
			\vspace*{-9pt}\\
		         &                   & Cu$(111)$  & Cu$(311)$            \\
			\vspace*{-9pt}\\
			\hline
			BLK-P& $E_{f}$           &  -285    &    -386            \\
			     & $E_{f-D2}$        &  -521    &    -653            \\
			     & $E_{f-TS}$        &  -548    &    -695            \\
			     & $E_{ads}$         &  -170    &    -271            \\
		     	 & $E_{ads-D2}$      &  -316    &    -448            \\
		         & $E_{ads-TS}$      &  -342    &    -489            \\
		         &  $\delta_{1}$ (\%)&  -2.8    &    -7.1            \\
		         &  $\delta_{2}$ (\%)&  -2.4    &    -2.6            \\

		    \hline
		    BL-P & $E_{f}$           &  -348    &    -338            \\
                 & $E_{f-D2}$        &  -632    &    -620            \\
                 & $E_{f-TS}$        &  -667    &    -658            \\
		         & $E_{ads}$         &  -235    &    -225            \\
		         & $E_{ads-D2}$      &  -449    &    -437            \\
		         & $E_{ads-TS}$      &  -485    &    -476            \\
		         & $\delta_{1}$ (\%) &  -1.6    &    +0.3            \\
		         & $\delta_{2}$ (\%) &  -1.3    &    -1.7            \\       		
		         		
		\end{tabular}
	\end{ruledtabular}
\end{table}

To match BLK-P on a finite Cu slab with a minimal strain [Fig.\ref{fig:1}(a)],  the supercell contains four BLK-P unit cells with the zigzag 
 direction aligned with five Cu$(311)$ surface unit cells along the step ($x$ direction), and four BLK-P unit cells with the armchair 
direction aligned with four Cu$(311)$ surface unit cells perpendicular to the step ($y$ direction). This corresponds  to $240$~Cu atoms in the Cu$(311)$ slab and to $N_P$$=$$64$ P atoms in the BLK-P included in the supercell, or a 4:5 ratio of P:Cu atoms at the interface (in the bottom BLK-P layer and the topmost metal layer). The interface structure of BLK-P/Cu$(311)$ shows that all Cu atoms at the Cu$(311)$ surface steps are shared between the two rows of BLK-P atoms and form a bond, while the Cu atom on the terrace do not form bonds to BLK-P atoms. This is because the lattice constant $a_{1(0)}^{BLK-P}$ of freestanding BLK-P is $7\%$ larger than the distance of $4.26$~\AA~ between steps on Cu$(311)$. 
The lattice mismatches are listed in Table \ref{tab:1} and defined as $\delta_{i}$= $(a_{i} - a_{i(0)})/a_{i(0)}$, where ${i}$ indexes the Cartesian components. Center-to center Cu-P bond length ranges from $2.25$~\AA~to $2.43$~\AA, with an average Cu-P bond length of $2.32$~\AA. However, the averaged interfacial {\it vertical} distance ($d$) between the bottom of BLK-P and the top most Cu layer is $1.68$~\AA. The average bond length above is close to the sum of covalent radii of Cu ($1.27$~\AA) and P ($1.08$~\AA) \cite{Zhu2014}, {\it i.e.}, $2.35$~\AA. This indicates the possibility of a considerable orbital overlap between P and Cu orbitals leading to chemical bonding. 

In comparison to BLK-P, BL-P exhibits a different adsorption pattern on Cu$(311)$ [Fig.\ref{fig:1}(b)]. In this case, even though all the Cu step atoms form bonds with a single row of BL-P atoms, every other Cu atom on the terrace also from bonds to some of the BL-P atoms. The supercell contains four BL-P unit cells with the zigzag
direction aligned with five Cu$(311)$ surface unit cells along the step, and three BL-P unit cells along the zigzag 
direction aligned with four Cu$(311)$ surface unit cells perpendicular to the step. The supercell includes $240$~Cu atoms and $N_P$$=$$48$~P atoms, or a 3:5 ratio of P:Cu atoms at the BL-P/metal interface.  The average lattice constants of BL-P are slightly increased to $a_{1}^{BL-P}$$=$$5.68$~\AA~($0.3\%$ increase) to match the Cu unit cell along the $y$ direction, and decreased to $a_{2}^{BL-P}$$=$$3.22$~\AA~($1.7\%$ reduction) along the $x$ direction. Center-to center Cu-P bond length ranges from $2.26$~\AA~ to $2.50$~\AA, with an average Cu-P bond length of $2.33$~\AA. The optimized interfacial distance ($d$) between the bottom of BL-P and the top most Cu layer is found to be $1.90$~\AA, which is $0.22$~\AA~ greater than BLK-P on Cu$(311)$.  

A comparison between phosphorous allotropes bonding on metal substrate is based on the formation energy $E_f$. This is defined with respect to a common reference state for the energies of P atoms, chosen here to be the tetraphosphorus P$_{4}$ molecule:
\begin{equation}
E_{f} = (E_{slab+P} - E_{slab} - N_{P}\times E_{P_{4}}/4)/N_{P} \,\,\,.  
\label{equ:Ef}
\end{equation}
Here,  $E_{slab+P}$, $E_{slab}$, and $E_{P_{4}}$ are the total energies of the slab with phosphorene on it, of the metal slab, and  of P$_{4}$, respectively. This picture assumes that the allotropes are assembled from fragments of the P$_4$ precursor dissociated on the surface. Another comparison can be made in terms of the adsorption energy $E_{ads}$, often used in the literature:
\begin{equation}
E_{ads} = (E_{slab+P} - E_{slab} - E_{P})/N_{P} \,\,\,,
\label{equ:Eads}
\end{equation}
where  $E_{P}$ is the energy of the isolated phosphorene allotrope. This assumes an allotrope already formed before it is adsorbed on the surface, such as in an exfoliation/transfer process. The calculated $E_f$ and $E_{ads}$ are summarized in Tab. \ref{tab:1}.
Negative values indicate bonding. Both BL-P and BLK-P bind to either surface, with BL-P bonding more strongly than BLK-P on Cu$(111)$, and BLK-P bonding more strongly that BL-P on Cu$(311)$. We note that the difference between $E_{ads}^{BLK-P}$ and $E_{ads}^{BL-P}$ follows closely the difference $E_{f}^{BLK-P}$$-$$E_{f}^{BL-P}$ for all chemistry levels considered. In addition, for both allotropes there is significant bonding already at the PBE level. The addition of dispersive force corrections further increase the bonding energies, maintaining the ordering BL-P, BLK-P. For BLK-P $E_{ads}$ on the Cu$(311)$ surface is relatively large, {\it i.e.}, $0.49$~eV. Using the $E_f$ criterion, we obtain that BLK-P is favored by $\approx$$40$~meV/atom on Cu$(311)$, whereas using $E_{ads}$ the difference is reduced to $\approx$$10$~meV/atom. The former criterion seems more realistic from the point of view of assembling the P allotropes from P fragments on the surface and indicates that BLK-P is stable  on Cu$(311)$ against transition to BL-P at room temperature.

\begin{figure}[th!]
	\centering
	\includegraphics[trim = 0.0cm 0cm 0cm 0cm, clip, scale=0.49, angle=0]{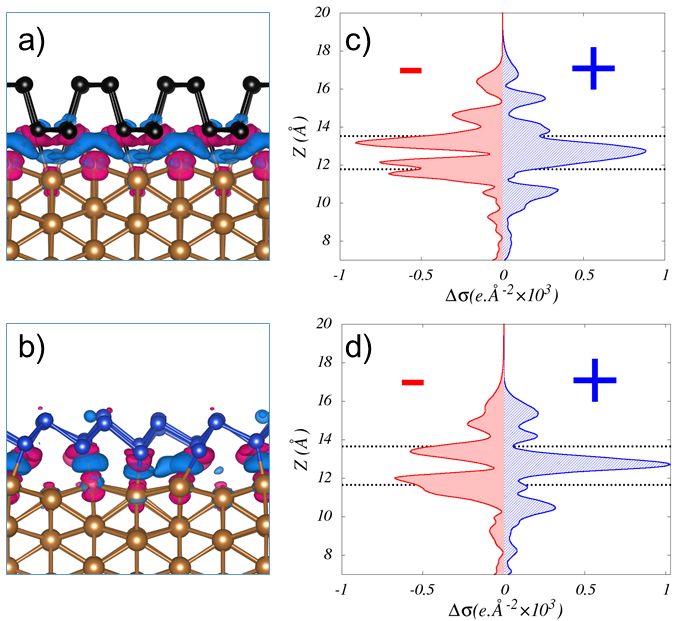}
	\caption{ \label{fig:2} Charge density difference plot of a) BLK-P and b) BL-P adsorbed on Cu$(311)$ surface, with the isosurface level of $0.03~e \times$\AA$^{-3}$ , where $e$ is the elementary electron charge. Blue and red regions denote loss (depletion) and gain (accumulation) of charges respectively. Variation in the averaged charge accumulation (blue) and depletion (red) planar density for (c) BLK-P/Cu$(311)$ and (d) BL-P/Cu$(311)$ respectively. The horizontal dotted lines in (c) and (d) represent the topmost Cu plane and the bottom P plane (see text for details).}
\end{figure}

An insight into the relation between geometries and adsorption energies can be obtained from the change in local electronic density at the interface defined as $\Delta \rho (\bm r)$$=$$\rho_{slab+P}$$-$$\rho_{slab}$$-$$\rho_{P}$. Figs. \ref{fig:2}(a),(b) show $ \Delta \rho (\bm r)$ for both allotropes using the same iso-values. The blue isosurfaces represent {\it local} charge migrating to new, hybrid orbitals formed by the P-Cu interaction. The red isosurfaces represent migration of electronic charge from the linear superposition of the separate Cu slab and P allotrope densities. In addition, Figs. \ref{fig:2}(c,d) show the plane average $\Delta \sigma_\pm(z)$$=$$\int \Delta \rho_\pm(\bm r)dxdy$ for charge migration to hybrid orbitals ($+$, blue) and depletion of the linearly-superposed orbitals ($-$, red). The topmost Cu plane and the lowest P plane are represented by dotted lines. We notice that $\Delta \sigma_-$ peaks on these planes and are separated by a charge minimum. On the other hand, $\Delta \sigma_+(z)$ is maximum at the interface between these planes and decays towards the two components of the system, with satellite minima on the second layer of Cu atoms and the topmost layer of P atoms. This is consistent with the $\Delta \rho$ iso-surface plots. We interpret this as an overall bonding between P and Cu atoms via hybrid states. A larger population in the hybrid orbitals is associated with stronger bonds, reflected in the adsorption energy. In the case of BLK-P, $\int \Delta \sigma_+(z)dz$ gives $0.166~e$ per surface Cu atom of the Cu$(311)$ substrate. For BL-P,  $\int \Delta \sigma_+(z)dz$ amounts to $0.146~e$ per Cu atom per surface Cu atom in Cu$(311)$. The net difference of $13.75$\% in charge transfer per surface Cu atom favors BLK-P.  On the other hand, in the case of the Cu(111) $\int \Delta \sigma_+(z)dz$ amounts to $0.135~e$ per surface Cu atom for BLK-P, and to $0.154~e$ per surface Cu atom for BL-P.  The net difference is $12.60$\% in favor of the latter. Besides the increased symmetry, a factor for the change in preference comes from the P:Cu ratios at the interface, {\it i.e.}, 4:5 and 16:25 for BLK-P and BL-P on Cu(111). This points to a slightly increased number of interface P-Cu bonds for BL-P on Cu(111) in comparison to Cu(311). These results are consistent with the ordering of adsorption energies.

\begin{figure}[th!]
\centering
\includegraphics[trim = 0.0cm 0.0cm 0.0cm 0.0cm, clip, scale=0.34, angle=0]{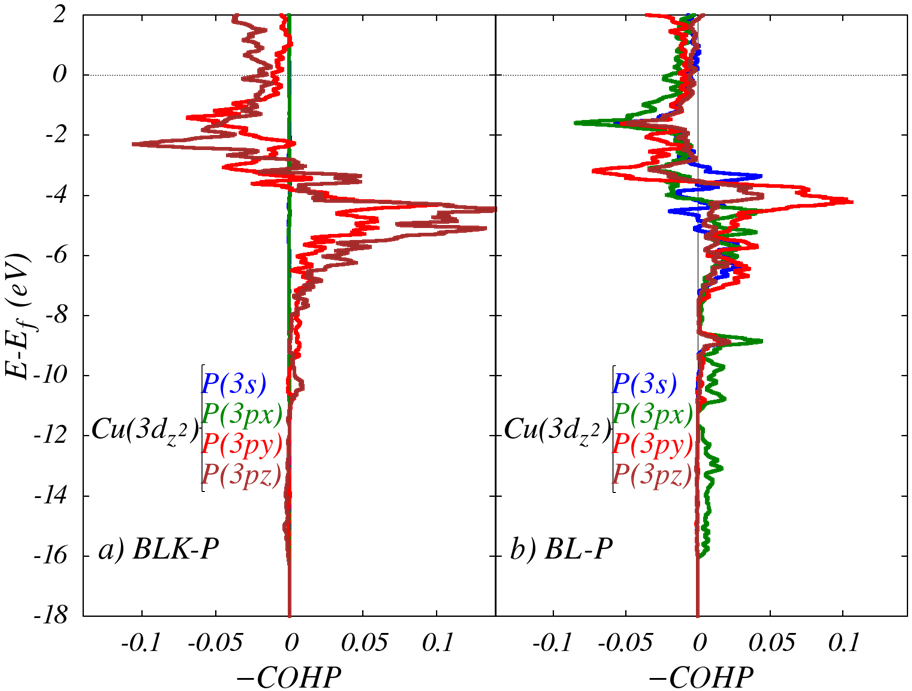}
\caption{ \label{fig:3} --COHP decomposition involving the Cu$_{d_{z^2}}$ and all P orbitals of (a) the BPK-P/Cu$(311)$, (b) the BL-P/Cu$(311)$. The various components show the dominating orbital interactions in each case.} 
\end{figure}

The discussion above on charge redistribution into hybrid states is a global view over all orbital types and energies. The nature of the hybrid states  can be understood by analyzing either the partial density of electronic states (PDOS) or by a Crystal Orbital Hamilton Populations (COHP) analysis \cite{Deringer2011,Dronskowski1993}. The latter emerged from the Frontier Orbital Theory \cite{Hoffman1998} and is directly related to the PDOS by multiplication with bond orbital populations extracted from the first-principles calculation. Two features usually occur upon absorption: a hybridization of metal Fermi level and the adsorbate conduction band minimum (or lowest unoccupied orbital), which give rise to an occupied state below the Fermi level and to an empty orbital higher in the metal conduction band; and a strong splitting of the deep adsorbate states. The latter usually results from a hybridization of the upper part of the adsorbate valence band (or highest occupied orbital) with the first unoccupied metal state. In our case, PDOS show that BLK-P undergoes a larger splitting of the deep P orbitals than BL-P. The magnitude of the splitting is associated with a larger bonding energy and is consistent with symmetry arguments. Namely, the laterally-extended $p$ orbitals of BLK-P have a $C_{2v}$ coordination \cite{Hoffman1998} that prefers the symmetry of the $(311)$ surface, whereas the vertically-extended $p$ orbitals of BL-P have a $C_{3v}$ symmetry  that prefer binding with vertical metal orbitals on Cu$(111)$. This is reflected in the COHP analysis which decomposes the bonding in shows the contributions from each orbital pair.

\begin{figure}[th!]
	\centering
	\includegraphics[trim = 0.0cm 0.0cm 0.0cm 0.0cm, clip, scale=0.34, angle=0]{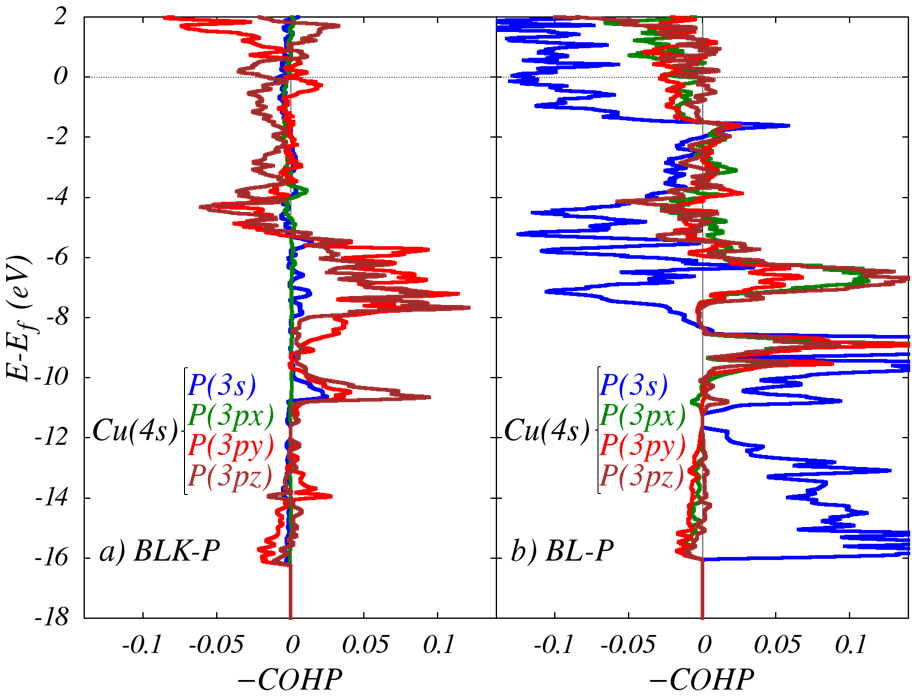}
	\caption{ \label{fig:4} --COHP decomposition involving the Cu$_s$ and all P dominant orbitals of (a) the BPK-P/Cu$(311)$, (b) the BL-P/Cu$(311)$.}
\end{figure}

Figs. \ref{fig:3},\ref{fig:4} show the COHP analyses on the $(311)$ surface. These are normalized by the number of bonds and are represented with a minus prefactor so that positive values of --COHP represent bonding, and negative values antibonding contributions. We show the contributions from the Cu$_{d_{z^2}}$ and Cu$_s$ only with all P orbitals, since we find that these have the largest --COHP peaks. In Fig. \ref{fig:3}, we notice strong bonding contributions from Cu$_{d_{z^2}}$ in the interval [$-6$,$-4$]~eV for both allotropes, but also large antibonding contributions in the interval [$-3$,$0$]~eV which cancel these bonding contributions. On the other hand, Fig. \ref{fig:4} shows  strong bonding contributions from Cu$_{s}$-P$_s$ in the interval [$-16$,$-8$]~eV for both allotropes, accompanied by much smaller Cu$_{s}$-P$_s$ antibonding contributions in the interval from [$-8$,$0$]~eV. In addition, the Cu$_{s}$-P$_p$ bonding contributions between [$-7$,$-4$]~eV for BLK-P, and [$-7$,$-6$]~eV for BL-P have very little cancellation from antibonding  Cu$_{s}$-P$_p$ close to the Fermi level. The larger footprint of Cu$_{s}$-P$_p$ bonding contributions noticed in the --COHP for BLK-P in comparison to BL-P, compounded with the larger number of P-Cu bonds described earlier, suggests that these are the bonds that favor the adsorption of BLK-P over BL-P on Cu(311). It is noticeable that in case of BLK-P the hybridization of P$_{p_x}$ orbitals is minimal and the bonding is dominated  by P$_{p_{y,z}}$ hybridization with Cu orbitals. This is expected from the orientation of P$_{p_y}$ orbitals (perpendicular to the step), in agreement with Ref. \cite{Hoffman1998}. In comparison, the BL-P $C_{3v}$ symmetry gives P$_p$ orbitals that are neither strictly orthogonal or parallel to the step, resulting in a non-vanishing Cu$_{s}$-P$_{p_x}$ contribution.

In conclusion, calculations within the DFT framework with vdW corrections suggest that BLK-P and BL-P  are selectively formed on Cu substrates, depending on the choice of low- and high-index Cu surfaces.  The high values of BLK-P and BL-P adsorption energies indicate strong interaction with the Cu substrates. The factors that favor BLK-P on stepped surfaces are the $C_{2v}$ symmetry of the surface and the increased DOS at the metal surface step. Here we used a simplified model system with a minimal terrace length, but we expect this to be valid for other metals and on wider terraces, albeit with different strain values. This work does not include alignment at different in-plane angles that may lead to Moire patterns. While the latter can lead to small variations in the adsorption energies on hexagonal and low-index surfaces, we expect that the $C_2v$ symmetry of BLK-P give significantly smaller adsorption energies for BLK-P unaligned with steps on large-index surfaces.

This work was supported in part by the DOD High Performance Computing Modernization Program and funded in part by the Air Force Office of Scientific Research under grant number FA9550-19RYCOR050. DH has been supported by an AFOSR-NRC postdoctoral fellowship through grant number FA9550-17-D-0001.

\bibliography{BLK-P_new}

\end{document}